# Development and Testing of a Low Cost Ultrasonic Leak Detector

Senol Gulgonul
Ostim Technical University
senol.gulgonul@ostimteknik.edu.tr

**Abstract.** This study focuses on the development of an ultrasonic leak detection system utilizing the Arduino Nano 33 BLE Sense Rev2 board. The research aimed to create a compact and cost-effective solution for identifying leaks in high-pressure pipes. Algorithms were designed to enable lossless recording and processing of sound data captured by the onboard MEMS microphone. Key signal processing techniques, including the implementation of an IIR high-pass filter and RMS calculation, were employed to detect ultrasonic frequencies associated with leaks. The system was tested on a pressurized pipe setup, demonstrating its ability to accurately identify leaks. The results highlight the system's effectiveness, with its compact design and low cost making it suitable for a wide range of industrial applications. This research contributes a practical and accessible tool for leak detection, offering potential benefits in industrial applications.

## 1. Introduction

Ultrasonic leak detection has become a critical technology for ensuring industrial safety and operational efficiency. By utilizing high-frequency sound waves beyond the range of human hearing, this non-destructive testing method can quickly and accurately identify gas leaks in pressurized systems. Early detection of leaks not only prevents potential hazards such as explosions and toxic gas exposure but also minimizes economic losses caused by wasted resources and unplanned downtime. Ultrasonic leak detection is versatile and applicable to a wide range of gases and environments, making it an indispensable tool in industries such as oil and gas, petrochemicals, and power generation. As industries continue to prioritize safety and sustainability, the importance of ultrasonic leak detection continues to grow.

When the pressure inside a vessel or pipe exceeds the external pressure and a leak hole is present, gas will escape. If the leak is small and the Reynolds number is high (Re > 4000), the gas outflow through the leak to the low-pressure side generates turbulence, which produces sound at specific frequencies. This turbulence generates ultrasonic sound. Typically, a smaller leak rate is associated with higher frequency sound [1].

Spectrum analysis plays a crucial role in ultrasonic leak detection, with frequency analysis forming the foundation of this process. Research indicates that the wave power spectrum is a function of the Strouhal Number [2]. Strouhal proposed an empirical equation that relates the frequency generated by turbulence to the velocity of air and diameter of hole, based on extensive experiments and analyses [3].

$$f = k\frac{v}{d} \quad (1)$$

where $f$ is the frequency of a sound wave, $v$ is the velocity of air in the leakage hole, $d$ is the average diameter of the hole, and $k$ is the proportional coefficient (in Hz/m²). In an experimental study of a pressurized vessel at 7 bar, an ultrasonic microphone placed 10 mm from the leak hole generated a peak in the ultrasonic spectrum around 40 kHz [4]. Parabolic or horn-type structures increase the sound level of the ultrasonic microphone, allowing measurements at a distance of 5 meters from the hole. A peak at 32 kHz was measured with the developed electronic leak detector board and parabolic structure [5].

Several commercial ultrasonic leak detectors use analog or digital circuits to detect ultrasonic sound levels above 20 kHz. For example, the Amprobe ULD-405 Ultrasonic Leak Detector operates in the 20 kHz to 90 kHz frequency range and features an LCD display to show the sound level in a bar graph [6]. The SKF Ultrasonic Leak Detector TKSU 10 has an LED display that provides guidance for the level of ultrasonic sound in the 35 to 42 kHz range [7]. In addition to visual displays, commercial leak detectors often include an audible buzzer sound, which can be used with a high-quality neckband headset for operation in noisy environments. Some leak detectors also feature a transmitter to detect leaks in unpressurized volumes, typically around the 40 kHz range [8].

## 2. Materials and Methods

The Arduino Nano 33 BLE Sense Rev2 features a built-in MEMS omnidirectional digital microphone (MP34DT06JTR), which captures sound using the Pulse Density Modulation (PDM) library. Its enhanced memory and processing power enable the digital analysis and filtering of captured sound data. The board is powered by the Nordic Semiconductor nRF52840, operating at a clock speed of 64 MHz. This microcontroller includes 256 KB of SRAM and 1 MB of flash memory. Additionally, the Arduino Nano 33 BLE Sense Rev2 is equipped with several built-in sensors, including:
- IMU: BMI270 (3-axis accelerometer + 3-axis gyroscope) + BMM150 (3-axis Magnetometer)
- Microphone: MP34DT06JTR
- Gesture, light, procimity: APDS9960
- Barometric pressure: LDS22HB
- Temperature, humidity: HS3003

The MP34DT06JTR is a low-power, omnidirectional digital MEMS microphone built with a capacitive sensing element and an integrated circuit (IC) that provides PDM output. It features low distortion, a 64 dB signal-to-noise ratio, and a sensitivity of -26 dBFS ±1 dB. The MP34DT06JTR is recommended for use in the acoustic range and has an almost flat frequency response from 100 Hz to 10 kHz [10]. Above 10 kHz, the gain starts to increase, and the MEMS membrane's mechanical resonance frequency is generally around 30 kHz. While this is not ideal for sound recording, it is useful for detecting ultrasonic sounds.

The Arduino PDM library enables the use of PDM microphones, such as the MP34DT06JTR. The original Arduino PDM library supports limited sampling rates (16,000 and 41,667) and buffer sizes. Therefore, a modified PDM library with a higher sampling rate of up to 62,500 and an increased buffer size was utilized in this study [11].

A Rossmax NA100 Ultra Compact Piston Nebulizer was used to create a pressurized plastic pipe [12]. The Rossmax NA100 features a powerful piston compressor with a pressure of 2.96 bar. One end of the pipe was connected to the nebulizer, while the other end was sealed using patafix glue, as shown in Fig. 1. The needle shown in Fig. 1 was removed to allow a leak for testing.

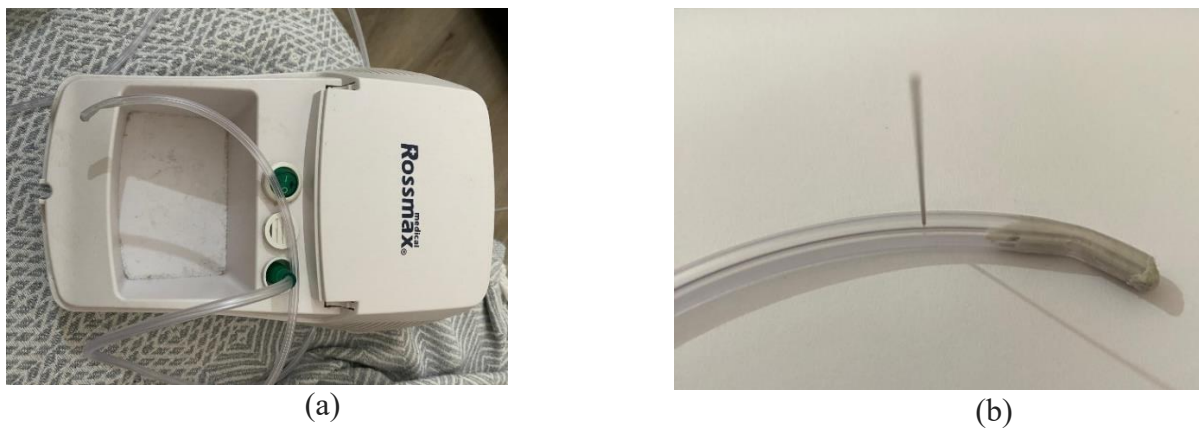

(a)  (b)

**Figure 1**. *Rossmax NA100 Nebulizer (a) and leak hole (b)*

There is no specific setup required for the Arduino Nano 33 BLE Sense Rev2. The Serial Plotter in the Arduino IDE is insufficient for time axis measurements and scaling. Therefore, we used the Better Serial Plotter, which provides a time axis, scalable y-axis, and multi-plot options. Additionally, the Better Serial Plotter can export data to .CSV format, facilitating the creation of measurement graphics in Microsoft Excel [13].

We developed two Arduino codes. The first code samples sound at a rate of 62,500. The received sound data is first recorded to a read buffer with a size of 62,500 and then transmitted via high-speed UART at a baud rate of 921,600. This method ensures that raw sound data is transmitted to the UART port quickly without data loss. A Python code running on a notebook reads the UART data from the Arduino Nano 33 BLE Sense Rev2 and converts it into a .wav file to analyze the sound captured by the Arduino MEMS microphone. The Python code reads data in chunks of 62,500 samples, writes it to a buffer, and then transfers the buffer content to a .wav file while reading the next chunk. This approach allows the recorded sound to be viewed in both the time and frequency domains using Audacity audio editor and recorder software.

When the nebulizer is turned on, it produces significant noise at low frequencies, as shown in Fig. 2. When the leak point is brought within 1 cm of the microphone, the audible sound does not change much, but the ultrasonic sound levels rise above 20 kHz, with a peak at 26 kHz. This setup test demonstrated that the leak generates ultrasonic sound. This was an interim step in our development process. The ultimate goal is to detect ultrasonic sounds and provide a level or alert signal to the user using only the Arduino.

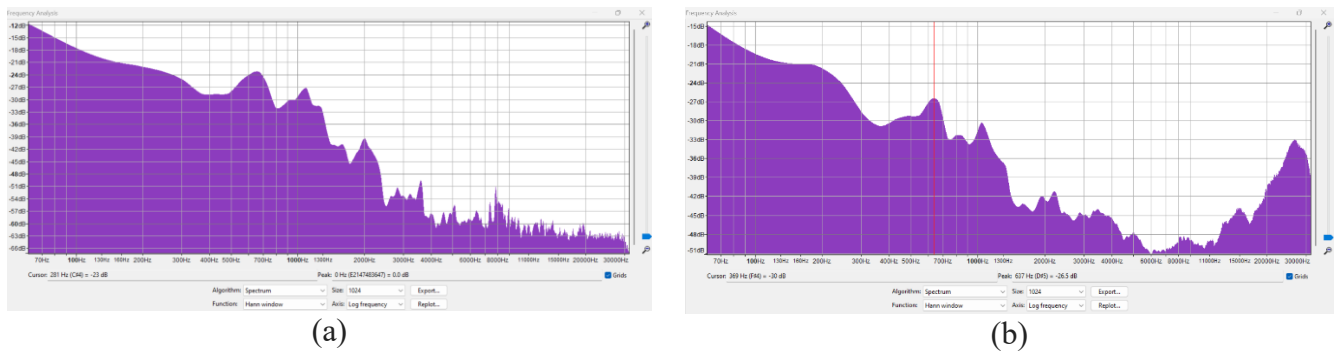

(a)                  (b)

**Figure 2**. *Spectrum with nebulizer working: leak is 1m far from microphone (a), leak is 1cm close to microphone (b)*

In the second code running on the Arduino Nano 33 BLE Sense Rev2, sound reading was performed using the tweaked PDM library at a 62,500 sample rate. Since we are only interested in the ultrasonic part of the sound, we applied a high-pass filter above 20 kHz and calculated the Root-Mean-Square (RMS) value to indicate the level of the ultrasonic sound. The high-pass filter implemented in this code is based on a first-order difference equation, which effectively attenuates low-frequency components while allowing high-frequency signals to pass.

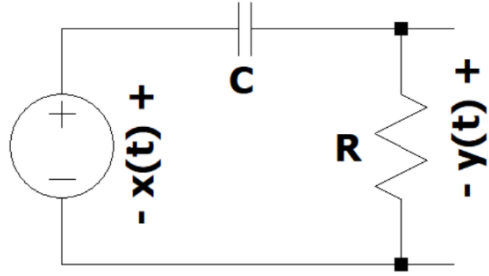

**Figure 3**. *First order analog high pass filter*

The equation of the first order analog high pass RC filter shown in Fig. 3 can be written:

$$y(t) = RC\left(\frac{dx(t)}{dt} - \frac{dy(t)}{dt}\right) \quad (1)$$

We can convert this differential equation to a difference equation:

$$y[n] = RC\left(\frac{x[n] - x[n-1]}{\Delta t} - \frac{y[n] - y[n-1]}{\Delta t}\right)$$

$$y[n] = \frac{RC}{RC + \Delta t}(y[n-1]) + x[n] - x[n-1]$$

$$y[n] = \alpha \cdot (y[n-1] + x[n] - x[n-1]) \quad (2)$$

where $y[n]$ is the current filtered output, $y[n-1]$ is the previous filtered output, $x[n]$ is the current input sample, and $x[n-1]$ is the previous input sample. The coefficient $\alpha$ determines the cutoff frequency of the filter and is calculated based on the desired cutoff frequency and the sampling rate. The equation for $\alpha$ is given by:

$$\alpha = \frac{2\pi f_c}{2\pi f_c + f_s} \quad (3)$$

where $f_c$ is the cutoff frequency and $f_s$ is the sampling rate.

This is actually a first order IIR filter with M=1, N=1, in general form [14]

$$y[n] = -\sum_{k=1}^{N} a_k y[n-k] + \sum_{k=0}^{M} b_k x[n-k] \quad (4)$$

In this implementation, α is set to 0.33, corresponding to a cutoff frequency of 20 kHz. This high-pass filter is applied to each sample in the buffer, effectively removing low-frequency noise and DC offset, thus preserving the integrity of the high-frequency components of the signal. The Arduino implementation of the high-pass filter and RMS calculation is shown in Fig.4. The calculated RMS results are printed to the Arduino UART port every 200 ms using the millis() function. The user can either read the RMS values directly or visualize them using the Better Serial Plotter.

```
/**
 * Function to calculate the RMS value of the high-pass filtered samples.
 */
float calculateRMS() {
  float rms = 0;
  for (int i = 0; i < samplesRead; i++) {
    // Apply high-pass filter
    filteredSample = alpha * (filteredSample + sampleBuffer[i] - previousSample);
    previousSample = sampleBuffer[i];

    // Accumulate the squared value of the filtered sample
    rms += filteredSample * filteredSample;
  }
  rms = sqrt(rms / samplesRead);
  return rms;
}
```

**Figure 4**. *Arduino implementation of high pass IIR filter and RMS calculation*

## 3. Results and Discussion

The RMS recording started with a period of silence, during which the RMS values were around 10. When the nebulizer was turned on, it produced audible noise, raising the RMS values to 50, but no ultrasonic spectrum was detected, as shown in Fig. 5. In the second phase, the leak point was approximately 1 meter away from the sensor. When the leak point was moved to within 1 cm of the sensor, a significant increase in RMS values was clearly visible in the graph.

It was not possible to distinguish a sound originating from a leak within the noise of the nebulizer. Therefore, we could not determine whether there was a leak by the sound heard with our ears. This situation is similar in noisy environments such as factories. While a leak may be noticeable from harmonics in the audible spectrum in a very quiet environment, it is impossible to distinguish in a noisy environment. For this reason, ultrasonic leak detectors are necessary.

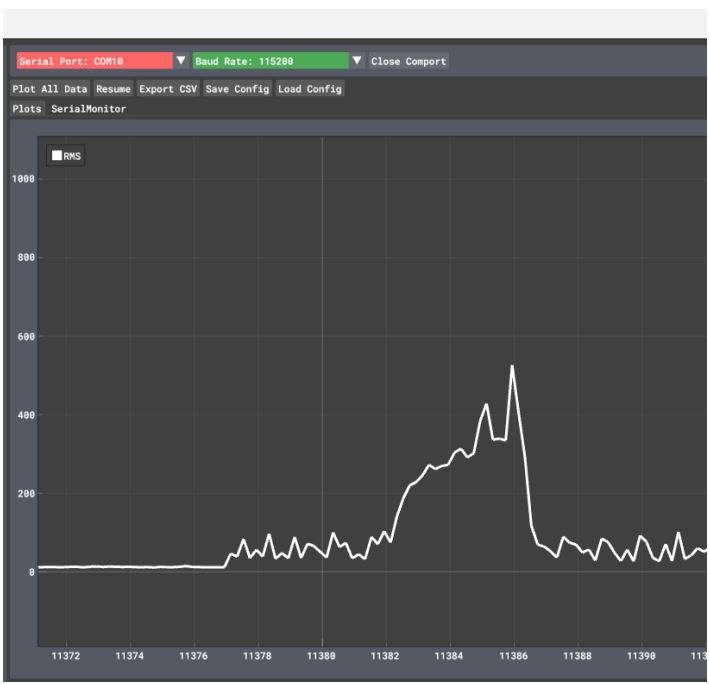

**Figure 5**. *RMS values for silence, nebulizer turned on and leak point close to sensor phases*

## 4. Conclusion

An Arduino Nano 33 BLE Sense Rev2 was programmed to detect ultrasonic leaks. Special algorithms were applied to buffer sound samples without causing any loss. The implemented first-order high-pass filter removed the audible sound spectrum and allowed the measurement of the RMS level of the ultrasonic sound spectrum. The prototype development proved the concept and provided insights into how ultrasonic leak detection can be realized.